# Engineering of Low-Loss Metal for Nanoplasmonic and Metamaterials Applications

D. A. Bobb, G. Zhu, M. Mayy, A. V. Gavrilenko, P. Mead, V. I. Gavrilenko, M. A. Noginov

*Center for Materials Research, Norfolk State University, Norfolk, VA, 23504*

**Abstract**

We have shown that alloying a noble metal (gold) with another metal (cadmium), which can contribute two electrons per atom to a free electron gas, can significantly improve the metal's optical properties in certain wavelength ranges and make them worse in the other parts of the spectrum. In particular, in the gold-cadmium alloy we have demonstrated a significant expansion of the spectral range of metallic reflectance to shorter wavelengths. The experimental results and the predictions of the first principles theory demonstrate an opportunity for the improvement and optimization of low-loss metals for nanoplasmonic and metamaterials applications.

Plasmonic materials and devices as well as metamaterials based on metallic nanoinclusions play an increasingly important role in modern photonics and nano-optics [1,2]. Their applications include but are not limited to surface enhanced Raman scattering (SERS) [3,4], scanning microscopy and spectroscopy with nanometer resolution [5,6], negative index of refraction [7,8], super-lens [9] and hyper-lens [10,11] with sub-diffraction resolution, transformation optics [12,13], and optical cloaking [14,15].

The common drawback of all metal-based metamaterials is high optical loss in metal at optical frequencies. It has been shown theoretically [16-18] and experimentally [19-22] that the optical loss in metallic nanostructures can be compensated and the stimulated emission of surface plasmons (SPs) can be obtained if the optical gain is added to an adjacent dielectric.

Conquering the optical loss by gain requires intense Q-switched laser pumping, which makes systems and devises prohibitively complex. In addition, some gain media (*e.g.* laser dyes) often have low resistance to photobleaching. This calls for alternative loss reduction techniques, which do not require laser pumping. It has been demonstrated [23] that the propagation length of surface plasmon polaritons (SPPs) can be significantly increased if the laser dye in very high concentration is added to the dielectric medium adjacent to the silver surface. This effect has been explained in terms of the modification of electronic surface states of silver in the presence of dye molecules.

Optical loss in noble metals in the visible and ultraviolet spectral ranges is determined by the combination of a free electron loss, defined by the damping constant in the Drude model, and the absorption loss due to the electron transitions between occupied bound *d* states and unoccupied hybridized *sp* states above the Fermi level [24]. It has been proposed that alloying a noble metal (which contributes one electron per atom to a free electron gas) with a small volume fraction of another metal (which contributes two or three electrons per atom to a free electron gas) can raise the Fermi level, shift the spectral position of the $d \rightarrow sp$ transition band, and, correspondingly, modify metal's reflection and absorption spectra [24,25], Fig. 1. It has been argued [24] that this effect is responsible for the yellowish coloration of brass, which is an alloy of reddish copper and silvery-white zinc.

If the fraction of the impurity metal (amount of n-doping) is sufficiently small, then the alloying will not modify the electron band dispersion of the noble metal significantly and its only effect will be the change of the Fermi level. However, at larger concentrations of the impurity metal, the change of the band structures of both $d$ and $s$ electrons can influence the electron energy spectrum much stronger.

According to Ref. [31], a blue shift of the spectral band associated with bound electrons can lead to a blue shift of the edge of metallic reflectance, Fig. 1b. An increase of the free electron density will increase the plasma frequency $\omega_p$ [24] and, correspondingly, shift the edge of metallic reflectance to shorter wavelengths even further. Combined with an increase of the damping constant (caused by electron scattering on impurity-induced defects), an increase of $\omega_p$ should increase the optical loss associated with free electrons. Thus, the improvement of plasmonic properties of noble metals *via* alloying can be achieved only as a result of careful composition optimization.

In this Letter, we report on the studies of spectroscopic properties of pure gold and gold-cadmium alloys. The electron configurations of Au ([Xe] $4f^{14}5d^{10}6s^{1}$) and Cd ([Kr] $4d^{10}5s^{2}$) suggest that cadmium can provide one extra electron to a free electron gas of the alloy, similarly to how zinc provides an extra electron to a free electron gas of brass.

The ground state and the electron energy structure of Au-Cd alloys has been studied by the Density Functional Theory (DFT), employing the CASTEP computational package [33] and the first principle pseudopotentials. The super-cell method has been used to model the Au-Cd alloys at cadmium concentration equal to 3.3 at.%. The reliability of this approach has been carefully proven by the detailed convergence test. We have found that few percent accuracy of the total energy convergence can be achieved at 5x5x5 **k**-point mesh and energy cut-off of 500 eV.

Optical functions of Au-Cd are calculated within the Random Phase Approximation (RPA) using as the inputs the eigenenergies and the eigenfunctions obtained from the ground state run [27]. We have used the equilibrium lattice constants for both bulk Au and Au-Cd alloy as calculated by the Local Density Approximation (LDA) method. The calculated value of the lattice constant of gold, $a$=4.013 Å, is somewhat smaller than the experimental value, a=4.08 Å [28]. The resulting compression of the lattice, effectively modifying the widths of the electronic bands, acts as a quasi-particle (QP) correction. No additional QP corrections have been applied to the electron energy structure in order to avoid substantial complexity of the calculations. Consequently, the positions of the energy bands are calculated with a relatively small systematic error, which effect can be eliminated if *relative* changes of the band structure are compared with the experiment.

The calculated optical spectra of pure gold are depicted in Fig. 2a. Four peaks (I–IV), corresponding to the transitions between the *d* band and the hybridized *s*-*p* band, can be clearly seen in the reflectance spectrum as well as in the spectrum of the imaginary part of the dielectric constant $\varepsilon''$. The spectrum of $\varepsilon''$ in the Au(96.7 at.%):Cd(3.3 at.%) alloy (along with that of pure gold) is shown in Fig. 2b. In the alloy, bands II–IV tend to shift to lower energies while band I shifts to higher energies and gets partly overlapped with band II.

The analysis of the calculated projected density of states indicates that the predicted modifications of the optical spectra of the alloy are caused by substantial redistributions of the oscillator strengths due to the additional resonances between *d*- and *s*-electron states (both occupied and unoccupied) of impurity and host atoms.

The gold and the gold-cadmium alloy (with nominal concentrations Au(90at.%):Cd(10at.%)) have been acquired in a form of ~20 $mm^3$ beads from Kurt J. Lesker Co. Two types of experiments have been conducted. In the first set of experiments, metallic beads were grinded and polished, and the reflectance spectra of shiny flat metallic surfaces were taken using a spectrophotometer equipped with an integrating sphere. Pure gold was more intense-yellow than a noticeably whiter gold-cadmium alloy. Polished alloy samples had darker islands (with sharp boundaries) embedded in a lighter metallic host. We assumed that the darker inclusions had smaller concentration of cadmium than the lighter surrounding.

In the second type of experiments, beads were placed in water and exposed to short-pulsed laser radiation ($t_{pulse}$≈10 ns, $\lambda$=532 nm, ~10 $J/cm^2$). The suspensions of nanoparticles have been produced *via* photo-ablation and their extinction spectra have been studied in a regular spectrophotometer setup.

The reflectance spectrum of pure gold and the absorbance spectrum of gold nanoparticles are plotted in Fig. 3 along with the spectra of real ($\varepsilon'_{\mathrm{Au}}$) and imaginary ($\varepsilon''_{\mathrm{Au}}$) parts of the dielectric function of Au [29]. The surface plasmon (SP) band of Au nanoparticles has a maximum at ~18900 $cm^{-1}$ (529 nm), at which $\varepsilon'_{\mathrm{Au}}$ is approximately equal to $-2\varepsilon'_{\mathrm{water}}$ (here $\varepsilon'_{\mathrm{water}}$ is the real part of the dielectric constant of water). This as well as a nearly symmetric shape of the SP band, indicate that metallic nanoparticles contributing to the SP absorption are sufficiently small (dipole approximation) and the contribution of larger nanoparticles is insignificant. The edge of the metallic reflectance range (~20700 cm$^{-1}$, 483 nm) corresponds to the energy at which $\varepsilon'_{Au}$ spectrum starts deviating from the Drude behavior.

The two maxima, which can be associated with bands I and II in Fig. 2, are seen in both the reflectance spectrum and the spectrum of $\varepsilon''_{Au}$. The same two bands, although slightly shifted are found in the absorbance spectrum of Au nanoparticles. The relatively small mismatch between the energy positions of the experimentally measured and theoretically calculated maxima [30] is due to the LDA-predicted compressed lattice, which effectively acts as a QP correction.

As follows from Fig. 4, depicting the reflectance spectra of pure gold, darker Au-Cd alloy (#1) and lighter Au-Cd alloy (#2), the addition of cadmium shifts the edge of the metallic reflectance range to higher energies (shorter wavelengths) by ~2900 cm$^{-1}$ (~66 nm). The reflectance of alloys in the visible and near-infrared ranges of the spectrum is lower than that of pure gold. This suggests that the imaginary part of the dielectric function $\varepsilon''$ (optical loss) is getting larger when gold is alloyed with cadmium. Both the significant expansion of the reflectance spectrum and the increase of the optical loss in the visible and near-infrared ranges of the spectrum are in accord with the predictions of the heuristic model [24]. With the increase of the concentration of cadmium, band I shifts to the higher energies and band II shifts to the lower energies – in excellent agreement with the results of the first principle calculations, Fig. 2.

Absorbance spectra of the suspensions of Au and Au-Cd nanoparticles feature the peak associated with the SP resonance as well as the wing of a much stronger high-energy band, which is due to the contributions of bound electrons, Fig. 4b. The SP band of the alloy (19120 cm$^{-1}$, 523 nm) is slightly blue shifted relative to that of pure gold (18900 cm$^{-1}$, 529 nm). This effect is in line with the increase of the concentration of free electrons and the corresponding increase of $\omega_p$. The SP band of the alloy is nearly symmetrical (after the background is subtracted) and significantly broader than that in pure gold. The symmetric character of the band suggests that the SP absorption is predominated by small nanoparticles (dipole approximation) and that the increase of its width is due to the increase of $\varepsilon''$ (loss) rather than the dispersion of particle sizes. This conclusion is in line with that drawn based of the analysis of the reflectance spectra. In the alloy, absorption band II is red shifted relative to that in pure gold, in agreement with the theoretical prediction (Fig. 2b) and the reflectance experiment (Fig. 4a), and absorption band I is practically not seen.

To summarize, we have demonstrated that alloying gold with cadmium (i) significantly expands the spectral range of metallic reflection, moving its high-energy edge from 19600 $cm^{-1}$ (510 nm) to 22500 $cm^{-1}$ (444 nm), (ii) causes a blue shift of the SP spectral band in metallic nanoparticles, from 18900 $cm^{-1}$ (529 nm) to 19120 $cm^{-1}$ (523 nm), (iii) increases the width of the SP spectral band and slightly lowers the reflectance in the visible and near-infrared ranges of the spectrum, and (iv) causes the shift of the bands in the bound-electron absorption spectrum, in a good agreement with the predictions of the first-principle theory. The first two results are in a good agreement with the predictions of a heuristic model [24], according to which *n* doping of gold (alloying Au with Cd) should lead to an increase of the Fermi level and the corresponding blue shift of both the SP resonance and the edge of the metallic reflectance spectrum. It has been shown both theoretically and experimentally that impurity concentrations ≥3 at.% are too high to keep the electron band dispersion unaffected. This is evidenced by the shifts of the maxima in the bound-state electron spectra depicted in Figs. 2b and 4a,b. The experimental Fig. 4b as well as relevant theoretical calculations do not show the anticipated blue shift [24] of the bound-electron band. At the same time, the calculations do show the predicted increase of the Fermi level. Unexpectedly, the electronic *d* band in the Au-Cd alloy has moved, closely following the change of the Fermi level, keeping the position of the absorption band intact. Furthermore, the alloying has increased the loss in the green-to-near-infrared range of the spectrum.

We conclude that *n* doping of a noble metal can significantly improve its optical properties in certain spectral ranges and make them worse in the other parts of the spectrum. This opens an opportunity for the improvement and optimization of metals for nanoplasmonic and metamaterials applications and gives hope to many dream applications [1,2].

The work was supported by the NSF PREM grant # DMR 0611430, NSF NCN grant # EEC-0228390, and NASA URC grant # NCC3-1035.

Figure 1.

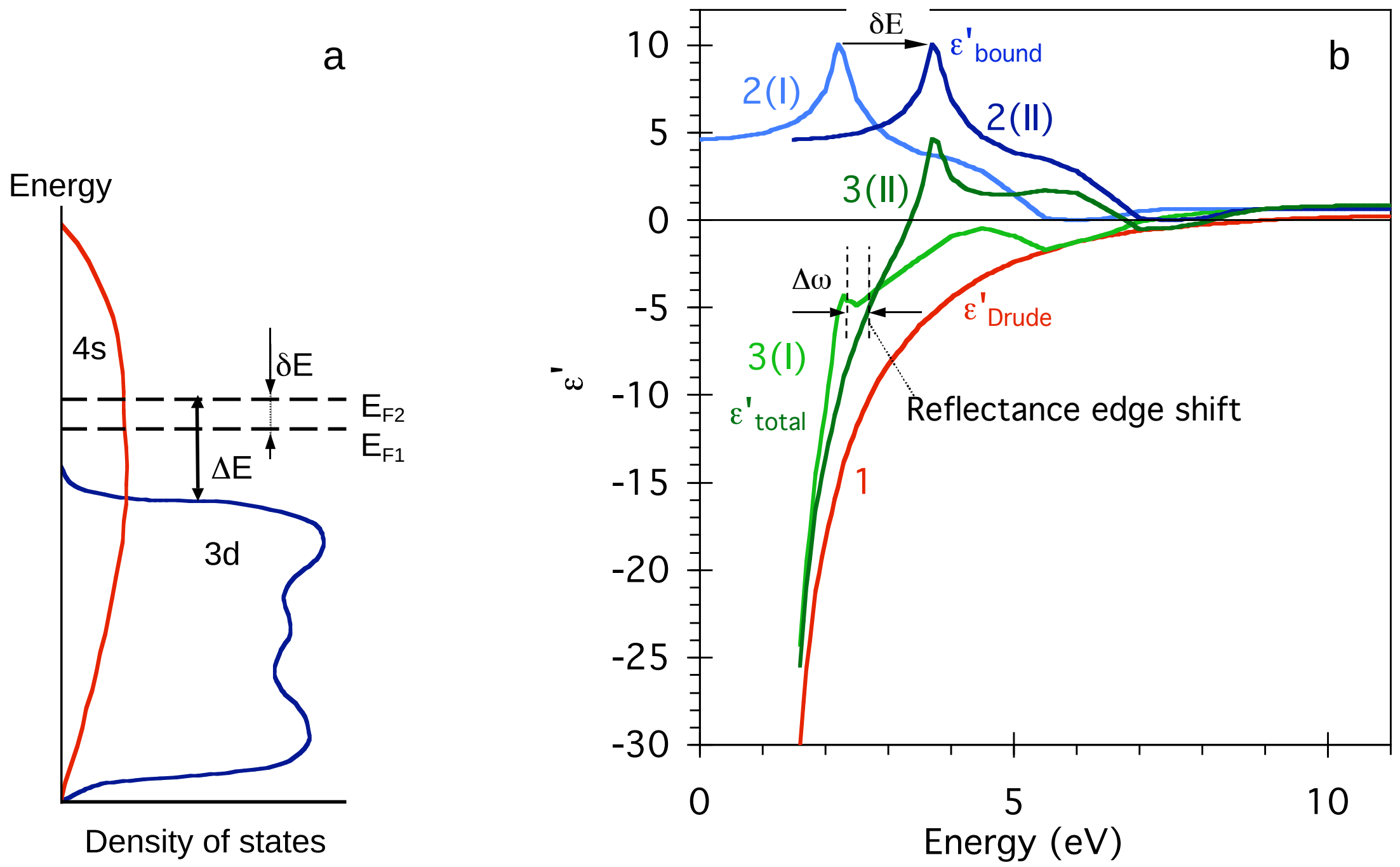

Fig. 1. (a) Schematic density-of-states diagram of a noble metal (copper), showing the bands of bound 3*d* electrons and free 4*s* electrons [24,25]. $E_{F1}$ and $E_{F2}$ are the Fermi levels in pure copper and copper alloyed with zinc; $\delta E=E_{F2}-E_{F1}$. The energy gap $\Delta E$ corresponds to the edge of the absorption band. (b) Spectrum of the real part of the dielectric function determined by the contributions of free electrons, $\varepsilon'_{\text{Drude}}$ (trace 1); spectra of the bound electron contributions, $\varepsilon'_{\text{bound}}$ (traces 2(I) and 2(II)); and net spectra $\varepsilon'_{\text{total}}$ calculated as the sums of free electron and bound electron contributions, (traces 3(I) and 3(II)). Spectra 2(I) and 3(I) correspond to pure copper and spectra 2(II) and 3(II) correspond to the alloy with an increased Fermi level. As a result of alloying, the edge of the spectral range of metallic reflectance is shifted by $\Delta\omega$.

Figure 2.

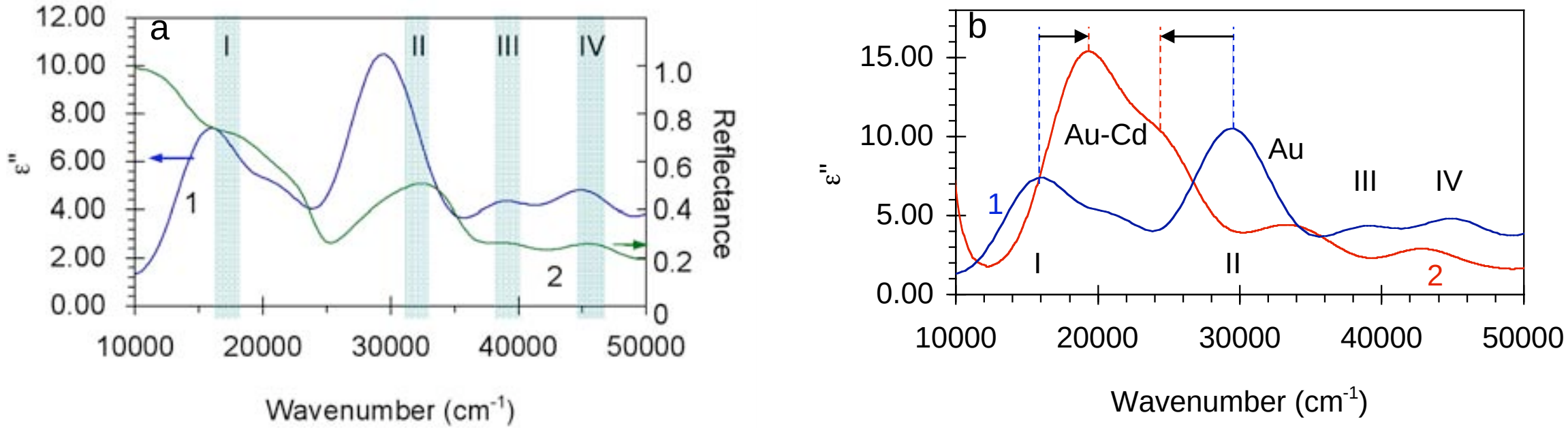

Fig. 2. (a) The spectrum of the imaginary part of the dielectric constant $\varepsilon''$ (trace 1) and the reflectance spectrum (trace 2) calculated for pure gold. (b) The spectrum of $\varepsilon''$ in pure gold (trace 1, same as in figure a) and Au(96.7at.%):Cd(3.3at.%) alloy (trace 2). The four characteristic peaks are denoted as I–IV.

Figure 3.

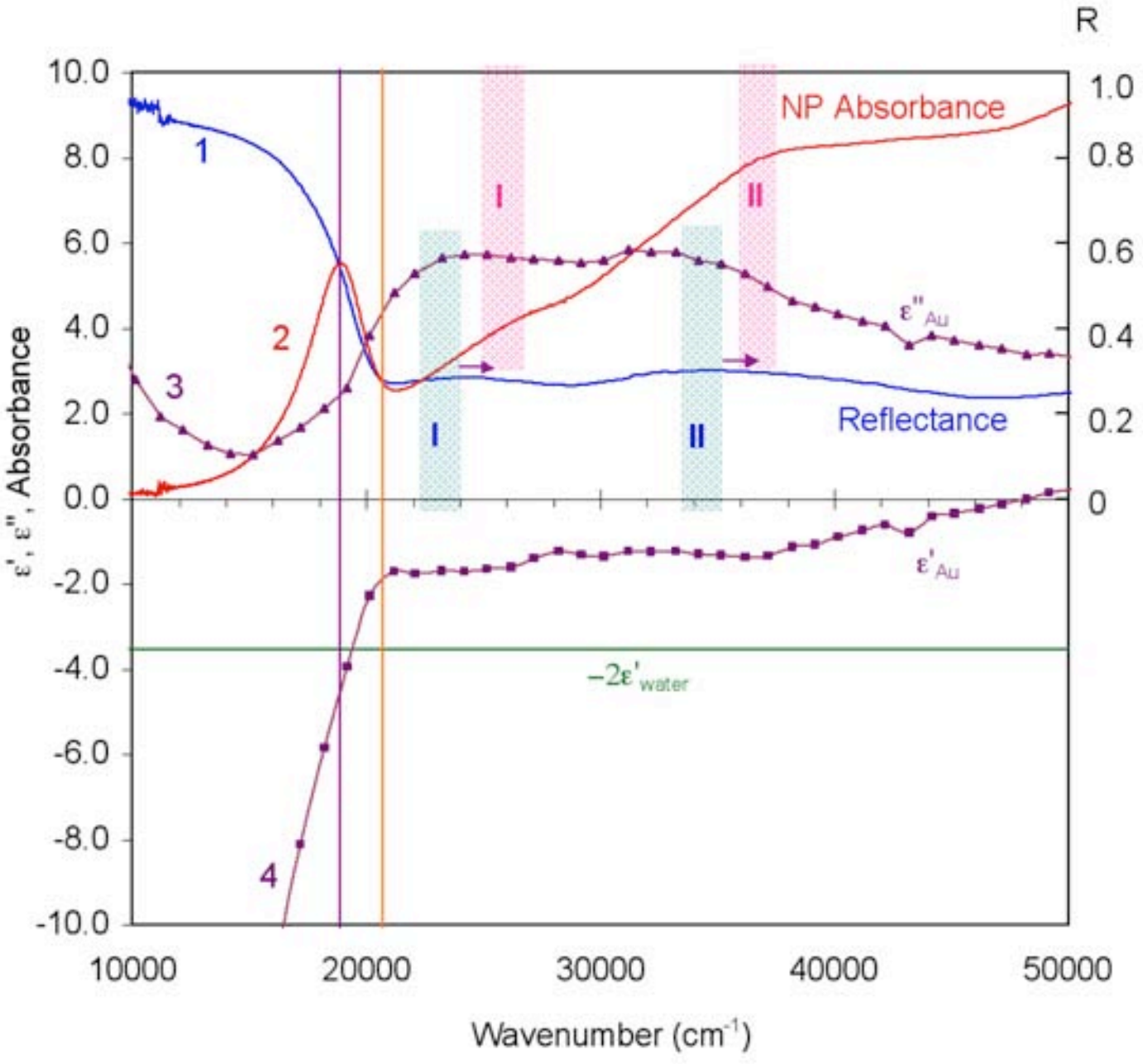

Fig. 3. Reflectance spectrum of gold (1), absorbance spectrum of the water suspension of gold nanoparticles (2), and the spectra of the real ($\varepsilon'_{\mathrm{Au}}$) and imaginary ( $\varepsilon''_{\mathrm{Au}}$) parts of the dielectric constants of gold [29]. Bands I and II are associated with those in Fig. 2. Horizontal line represents $-2\varepsilon'_{\mathrm{water}}$ and vertical lines are guides for eye.

Figure 4.

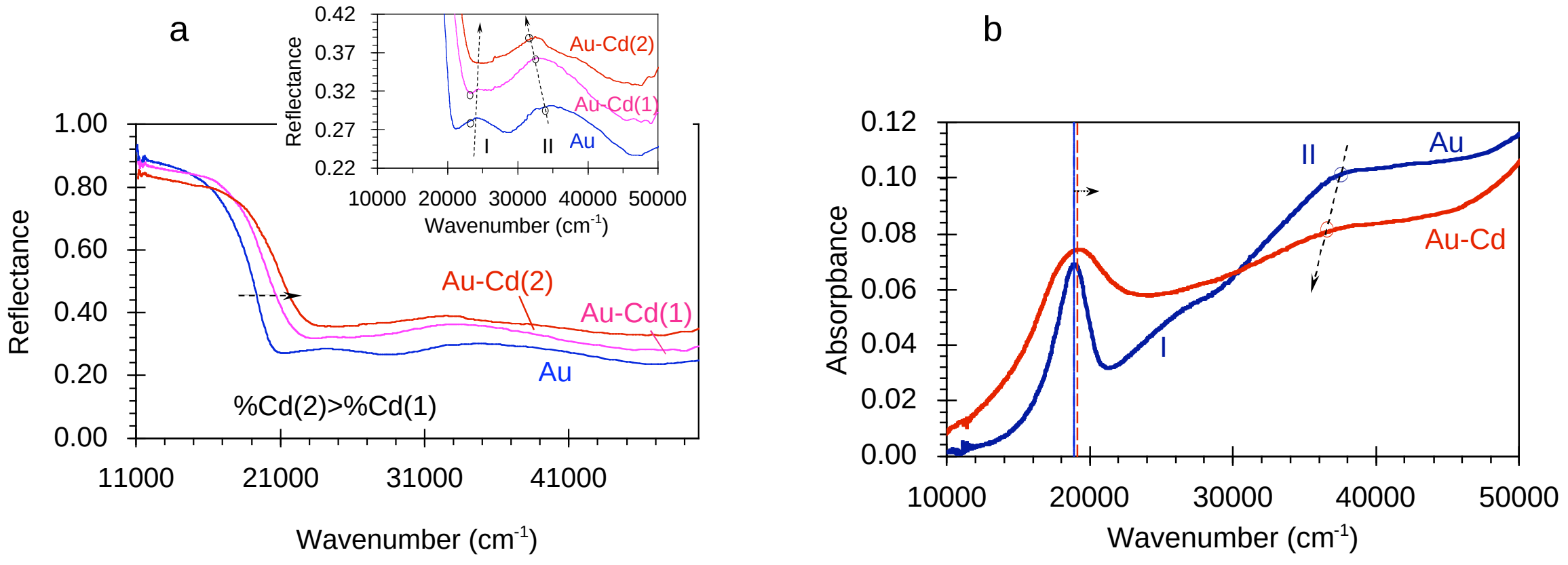

Fig. 4. (a) Absorption spectra of pure gold, darker Au-Cd alloy (#1) and lighter Au-Cd alloy (#2). Inset: zoomed parts of the same spectra showing the shifts of bands I and II. (b) Absorbance spectra of water suspensions of Au and Au-Cd nanoparticles. The arrows show the shifts of the spectral bands.